\documentclass[%
 reprint,
nofootinbib,
 amsmath,amssymb,
 aps,
 prd,
]{revtex4-2}
\usepackage{xcolor}
\usepackage{hhline}
\usepackage[usenames,dvipsnames,svgnames]{xcolor}
\usepackage[colorlinks=true,citecolor=purple,urlcolor=blue,linkcolor=olive]{hyperref}
\usepackage{graphicx}
\usepackage{dcolumn}
\usepackage{bm}
\usepackage{ulem}
\usepackage{array}
\usepackage{hyperref}
\usepackage{tabularray}
\UseTblrLibrary{booktabs}

\usepackage{multirow}
\usepackage[caption=false]{subfig}
\usepackage{natbib}
\begin{abstract}

We investigate, for the first time, universal relations for anisotropic dark energy stars. The stars are modeled with the modified Chaplygin equation of state and the Bowers–Liang prescription for anisotropy, and their global properties and $f$-mode frequencies are computed using the modified relativistic Hartle–Thorne slow rotation and Cowling approximations. We find that relations among moment of inertia, tidal deformability, quadrupole moment and $f$-mode frequency exhibit universality, with deviations limited to $\sim 1–10\%$, in close agreement with other compact star models.
Using tidal deformability constraints from GW170817 and GW190814, we obtain astrophysical limits on canonical properties of dark energy stars. 
For positive anisotropy strength, the radius of a $1.4M_\odot$ star is constrained to $R_{1.4}=8.93^{1.88}_{1.40}$ km (GW170817) and $10.92^{+0.71}_{-0.54}$ km (GW190814), consistent with observational bounds. The corresponding $f$-mode frequencies are constrained to $3.257^{+0.450}_{-0.537}$ kHz and $2.692^{+0.137}_{-0.157}$ kHz. 
Further, applying Pearson correlation analysis for the first time to anisotropic compact stars, we obtained the coefficients between various stellar attributes of dark energy stars and we show that the Chaplygin parameter $B$ correlates strongly with the $f$-mode frequency, with positive anisotropy strengthening while negative anisotropy weakening the correlation strength. These results establish  that universal relations extend to anisotropic dark energy stars and can be directly tested with present and future gravitational-wave observations. 
\end{abstract}

\begin{document}

\title{Universal Relations of Anisotropic Dark Energy Stars and Gravitational-Wave Constraints}

\author{O. P. Jyothilakshmi}
\email{op\_jyothilakshmi@cb.students.amrita.edu}
\author{V. Sreekanth}
\email{v\_sreekanth@cb.amrita.edu}

\affiliation{Department of Physics, Amrita School of Physical Sciences, Amrita Vishwa Vidyapeetham, Coimbatore, India}

\date{\today}
\maketitle

\section{Introduction}
The internal composition of compact stars remains unknown to this day, leading to the conjunctures of stars with varying constituents, such as normal nuclear matter, hyperons, quarks, Bose-Einstein condensates, dark matter, and dark energy. Despite this ambiguity in the stellar composition, some of the observational measurements for mass, radius, redshift and tidal deformability have introduced bounds, that help in constraining the stellar equation of state (EoS).  Observations from the gravitational wave (GW) events like GW170817 and GW190814, along with measurements of millisecond pulsars such as PSR J0348$-$0432~\cite{Antoniadis:2013pzd}, PSR J1614$-$2230~\cite{Fonseca:2016tux}, PSR J0740+6620~\cite{NANOGrav:2019jur}, and HESS J1731$-$347~\cite{Doroshenko2022} have significantly contributed to our understanding, in this regard. 

Although these observational constraints are valuable, they still carry considerable uncertainty for the constituent EoS. In this context, \textit{universal relations} become particularly important, as they provide nearly EoS-independent relations between the observable quantities like moment of inertia, tidal deformability, quadrupole moment, and mode frequencies of the star. These relations allow for a more effective probing of the fundamental properties of the compact stars and help us to tighten the constraints on the constituent EoS. 
Yagi and Yunes, first introduced a universal relation between the moment of inertia $I$, quadrupole moment $Q$ and tidal deformability $\Lambda$ of neutron stars~\cite{Yagi:2013awa}, which is the well known $I$-Love-$Q$ relation. The discovery of similar universal relations~\cite{Yagi:2013awa,Majumder:2015kfa,Steiner:2015aea,Chan:2015iou,Yagi:2016bkt,Bandyopadhyay:2017dvi,Mariji:2017buj,Wei:2018dyy,Kumar:2019xgp,Sotani:2021kiw,Largani:2021hjo,Das:2022ell,Zhao:2022tcw,Pradhan:2022rxs,Nath:2023gmu,Li:2023owg,Ghosh:2024cay} has recently provided new insights into observational astrophysics. 

Generally, astrophysical studies assume isotropic matter in the interior~\cite{Tolman:1939jz,Oppenheimer:1939ne,Cowling:1941nqk,Hartle:1967he,Hartle:1968si,Detweiler:1985zz,Lindblom:2001hd,Hinderer2008,Hind2010,Jha2008a,Jyothilakshmi:2022hys}. However, highly dense systems such as compact stars may exhibit anisotropy, i.e., unequal radial and tangential pressures. 
Various factors may contribute to this pressure anisotropy, such as strong magnetic field, superfluid core, rotation, relativistic interactions, and so on. In 1972, Ruderman introduced the idea of anisotropy in compact stars~\cite{Ruderman:1972aj}. Later, Bowers and Liang, introduced a model to study the effect of anisotropy in relativistic systems~\cite{Bowers:1974tgi}. They also derived the modified structural equations for a static anisotropic stellar configuration. The study of various attributes of the anisotropic compact stars has attracted great attention recently~\cite{Doneva:2012rd,Biswas:2019gkw,Pattersons:2021lci,Curi:2022nnt,Mondal:2023wwo,Lopes:2024wfk,Jyothilakshmi:2024zqn,Beltracchi:2024dfb,Jyothilakshmi:2024zso,Lau:2024oik}. Non-radial oscillations have been studied using the Cowling approximation~\cite{Cowling:1941nqk} prescription for anisotropic compact objects such as neutron stars~\cite{Doneva:2012rd}, hadronic and quark stars~\cite{Curi:2022nnt}, and dark energy stars~\cite{Jyothilakshmi:2024zqn}. 
Recently, fully relativistic treatments of the non-radial oscillations have also been explored in the case of anisotropic neutron stars~\cite{Mondal:2023wwo,Lau:2024oik}.
Also, the effect of anisotropy on the tidal deformability has been examined~\cite{Biswas:2019gkw} using methods similar to those employed for the isotropic case~\cite{Hinderer2008,Hind2010}.
The rotational properties of anisotropic compact stars have been studied in Refs.~\cite{Pattersons:2021lci,Beltracchi:2024dfb,Jyothilakshmi:2024zso} by modifying the Hartle-Thorne slow rotation approximation~\cite{Hartle:1967he,Hartle:1968si}.
Further, there has been a considerable interest to obtain the universal relations for anisotropic compact stars, in the recent times. The $I$-Love-$Q$ and other universal relations for anisotropic neutron and quark stars have been analyzed in Ref.~\cite{Yagi:2015hda}. Later, universal relations including the $f$-mode frequency of anisotropic neutron~\cite{Mohanty:2023hha} and interacting quark stars~\cite{Pretel:2024pem} have also been investigated. 

Among the many possible stellar models, compact objects made of elusive dark energy have attracted interest among the researchers for sometime~\cite{Lobo:2005uf,Chan:2008ui,Ghezzi:2009ct,Yazadjiev:2011sd,Rahaman:2011hd,Beltracchi:2018ait,Sakti:2021mvd,Abellan:2023tft,Panotopoulos:2021dtu,Panotopoulos:2020kgl,Pretel:2023nhf,Panotopoulos:2024iag,Jyothilakshmi:2024zqn,Jyothilakshmi:2024zso,BagheriTudeshki:2023dbm,Banerjee:2025zhp,Tangphati:2025fmm}. 
The idea that dark energy can be composed to form a compact star was first proposed by Chapline in 2004~\cite{Chapline:2004jfp}; and since then, there has been a significant interest in exploring the structure and characteristics of such compact objects. A widely used model to study the attributes of dark energy star is that of Chaplygin gas~\cite{Kamenshchik:2001cp,Bilic:2001cg,Bento:2002ps,Kahya:2015dpa}.
The global properties of slowly rotating isotropic dark energy stars under the extended Chaplygin EoS framework were also analyzed~\cite{Panotopoulos:2021dtu}. There are studies on the radial oscillations and tidal Love numbers of isotropic 
dark energy stars~\cite{Panotopoulos:2020kgl}. 
\par 
Lately, anisotropic dark energy stars have been looked at with enthusiasm. The rotational properties of anisotropic dark energy stars upto quadrupole deformation were obtained in Ref.~\cite{Jyothilakshmi:2024zso}; whereas the radial oscillations were studied in Ref.~\cite{Pretel:2023nhf,Panotopoulos:2024iag}. The non-radial oscillations of isotropic and anisotropic dark energy stars were investigated very recently~\cite{Jyothilakshmi:2024zqn} and the results showed a distinct oscillation spectra, which may facilitate the identification of such stars with the next generation GW detectors. 
\par 
In this Letter, we examine universal relations for anisotropic dark energy stars, focusing on stellar attributes such as moment of inertia, tidal deformability, quadrupole moment and $f$-mode frequencies. Using observational constraints from GW events, we derive theoretical limits on these quantities and perform Pearson correlation analysis to quantify their correlations systematically. 
We discuss the details of equation of state and formalism used in our work in Sec. \ref{sec:form}. Next, we discuss the results obtained in Sec. \ref{sec:result}. Finally, the summary is given in Sec. \ref{sec:summ}. 
\textit{Notations and conventions}: 
Newton's universal gravitational constant $G$ and velocity of light in vacuum $c$ are set to $G=c=1$, throughout this manuscript. M$_\odot$ represents  mass of the Sun. Metric convention followed is $g_{\mu\nu}=diag(-1,\,1,\,1,\,1)$. 


\section{Formalism}\label{sec:form}
\begin{figure}[h!]
    \centering
    \includegraphics[width=\linewidth]{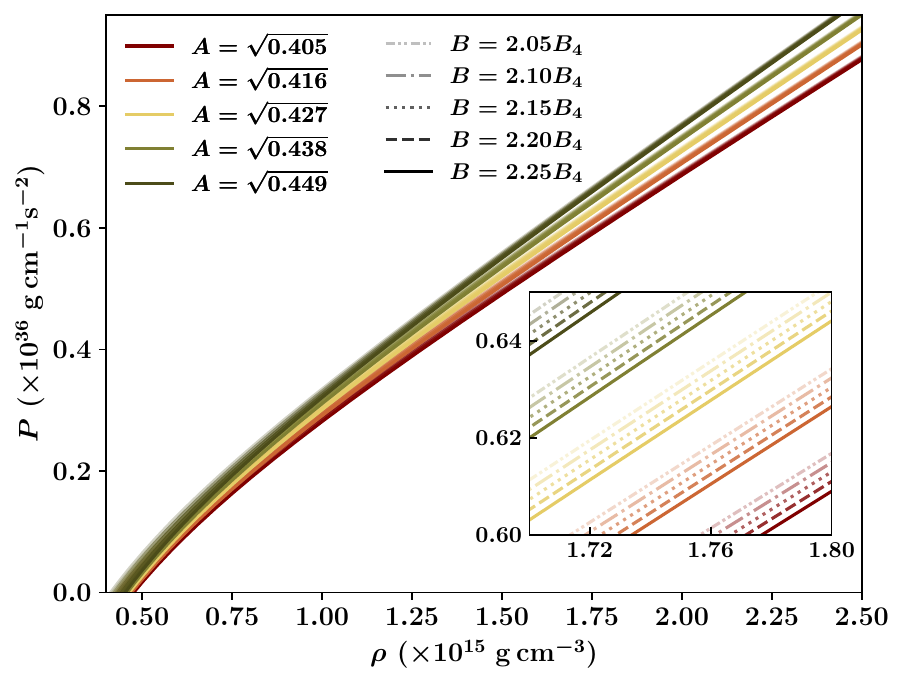}
    \caption{Pressure versus energy density plot for the Modified Chaplygin gas model by varying the parameters $A$ from $\sqrt{0.4}$ to $\sqrt{0.45}$ and $B$ 
    from $2.05B_4\, \textrm{to} \,2.25 B_4$ ($B_4 = 10^{-4}$ km$^{-2}$).
    }
    \label{fig:EoS}
\end{figure}

We study dark energy stars by adopting an equation of state (EoS) based on the modified Chaplygin gas model, a framework that unifies the behavior of dark matter and dark energy~\cite{Kamenshchik:2001cp,Bilic:2001cg,Bento:2002ps}. This model assumes that, the universe is filled with dark energy; and for our analysis, we employ the following EoS prescription~\cite{Kahya:2015dpa}: 
\begin{align}
    p = A^2 \rho - \frac{B^2}{\rho}; \label{EOS}
\end{align}
where, $p$ is the pressure and  $\rho$ is the energy density. The quantities $A$ (dimensionless) and $B$ (with units of energy density) are positive constants.
At the stellar surface, the pressure vanishes, leading to $\rho_s = B/A$. To maintain causality, the speed of sound must satisfy $v_s^2 = dp_r/d\rho < 1$, which implies the constraint: $A^2 < 0.5$~\cite{Pretel:2023nhf}. We construct a set of 25 pressure-density profiles, by varying the EoS parameters $A$ and $B$, within the ranges $\sqrt{0.4}$ to $\sqrt{0.45}$ and $0.2\times10^{-3}$ km$^{-2}$ to $0.23\times10^{-3}$ km$^{-2}$, respectively. The corresponding EoS curves are shown in Fig.~\ref{fig:EoS}.

Next, we proceed to study the various attributes of anisotropic dark energy stars, such as stellar mass, radius, tidal deformability, moment of inertia, quadrupole moment and $f$-mode frequency, under general relativistic treatment. 
The stress-energy tensor for an anisotropic fluid is given by
\begin{align}
    T^{\mu}_{\nu} = (\rho + q) u^\mu u_\nu + q \delta^{\mu}_{\nu} + \sigma k^{\mu}k_{\nu}; \label{Tmunu}
\end{align}
where, $\sigma = p - q$ is the anisotropy pressure, with $p$ and $q$ representing the radial and tangential pressures, respectively. Here, $u^\mu$ is the fluid four-velocity, and $k^\mu$ is a unit radial vector orthogonal to $u^\mu$; satisfying the conditions $u_\mu u^\mu = -1$, $k_\mu k^\mu = 1$, and $u^\mu k_\mu = 0$. 
To obtain the stellar structure of a static spherically symmetric system, we consider the metric;
\begin{align}
    ds^2 = -e^{2\nu} dt^2 + e^{\Lambda} dr^2 + r^2(d\theta^2 + \sin^2\theta d\phi^2),
\end{align}
where, $\nu(r)$ and $\Lambda(r)$ are the metric functions. 
By solving Einstein's field equations for the static, spherically symmetric metric with anisotropic fluid, one yields the modified form of Tolman-Oppenheimer-Volkoff (TOV) equations
~\cite{Tolman:1939jz,Oppenheimer:1939ne,Bowers:1974tgi}:
\begin{subequations}
\begin{align}
    \frac{dp}{dr} &= -\frac{(\rho+p)(m+4\pi r^3 p)}{r^2(1-2m/r)} - \frac{2\sigma}{r}, \\
    \frac{dm}{dr} &= 4\pi r^2 \rho.
\end{align}
\end{subequations}
Here, if we set $\sigma=0$, these equations reduce to that of isotropic scenario. 
We solve the above coupled differential equations from the center to the surface of the star, starting with the initial conditions $p(0) = p_c$ and $m(0) = 0$, with $p_c$ denoting the value of pressure at center($r=0$). The integration continues until the pressure vanishes at the stellar surface, i.e., $p(R) = 0$. The mass of the star is then given by $m(R) = M$, where $M$ is the total mass of the star.
We use the Bowers-Liang prescription to incorporate effects of anisotropy and in this model the anisotropy pressure takes the form \cite{Bowers:1974tgi}:
\begin{align}
    \sigma = -\lambda_{BL} \frac{r^3}{3}(\rho + 3p)\left(\frac{\rho + p}{r - 2m}\right);
\end{align}
where, $\lambda_{BL}$ measures the strength of anisotropy. We analyze the stellar structure for the following values of $\lambda_{BL} = -2,\, 0$ and $+2$.

Now, we use the pressure $p(r)$ and mass $m(r)$ profiles obtained for the static case to calculate the various global properties of the anisotropic dark energy stars. We follow Ref.~\cite{Jyothilakshmi:2024zso} to evaluate the rotational properties of anisotropic dark energy stars, such as the moment of inertia and quadrupole moment. 

The moment of inertia ($I=J/\Omega$) of a slowly rotating anisotropic star, with angular momentum $J$ and angular velocity $\Omega$, is given as~\cite{Rahmansyah:2020gar}
\begin{align}
    I = \frac{8\pi}{3} \int_0^R (\rho + p + \sigma) e^{\lambda - \nu} r^4 \left(\frac{\varpi}{\Omega}\right) dr;
\end{align}
where, 
$\varpi = \Omega - \omega'$ is its angular velocity relative to the local inertial frame.
We also calculate the mass quadrupole moment $Q$ of anisotropic dark energy stars~\cite{Hartle:1967he,Hartle:1968si} 
\begin{equation*}
    Q = \frac{8}{5}KM^3 + J^2/M;
\end{equation*}
with, $K$ being a constant obtained by solving the quadrupole deformation equations.

The tidal deformability $\Lambda$ quantifies the quadrupole deformation of a star under an external tidal field; and it is expressed as~\cite{Hinderer2008,Hind2010}:
\begin{align}
    \Lambda = \frac{2}{3} k_2 R^5.
\end{align}
Here, $k_2$ is the tidal Love number and $R$ is the radius. The Love number depends on the compactness $C=M/R$ and the function $y$ is defined in terms of the perturbation function $H$ as~\cite{Hinderer2008}
\begin{align}
    y=\frac{RH'(R)}{H(R)} - \frac{4\pi R^3 \varepsilon_s}{M(R)};
\end{align}
where, $\varepsilon_s$ being the surface energy density of the dark energy star. 
Furthermore, we obtain the $f$-mode frequencies of anisotropic dark energy stars, by employing the Cowling approximation~\cite{Cowling:1941nqk}, which has been successfully employed in the context of anisotropic dark energy stars~\cite{Jyothilakshmi:2024zqn}. 
Though this approximation is known to overestimate the results by $\sim 30\%$, it has been widely used in the studies of non-radial oscillations~\cite{Doneva:2012rd,Curi:2022nnt,Zhang:2023zth,Pretel:2024pem,Ghosh:2024cay,Jyothilakshmi:2024zqn,Jyothilakshmi:2024xtl,Jyothilakshmi:2025wru}. 
The mode frequency $f$ is obtained by solving the coupled differential equations given in Ref.~\cite{Doneva:2012rd} with appropriate boundary conditions. 

In order to study the universal relations, we normalize the quantities as:
\begin{align}
    \bar{I} = \frac{I}{M^3}, \qquad \bar{q} = \frac{Q}{M^3}, \qquad \omega = 2\pi f.
\end{align}
Thus, we obtain $\bar{I},\,{\Lambda},\,\bar{q},$ and $\omega$ for different stellar equilibria of dark energy stars needed for our analysis, by employing the set of EoS prescriptions (Fig.~\ref{fig:EoS}), for different values of anisotropy parameter $\lambda_{BL}=-2,\,0,$ and $+2$. 
\section{Results and Discussions}\label{sec:result}

In this section, we explore the various universal relations for anisotropic dark energy stars, focusing on the relationships between different attributes of compact stars, such as the radius $R$, mass $M$, moment of inertia $I$, tidal deformability $\Lambda$, quadrupole moment $Q$ and $f$-mode frequency. 
For this, we vary the model parameters $A$ and $B$ discussed in Sec.~\ref{sec:form} and generate 25 sets of EoSs. 
For each relation obtained, we also calculate the error or deviation from the numerical results using the expression:
$\Delta \eta = |\eta_{\text{fit}} - \eta|/\eta_{\text{fit}}$;
where $\eta$ is the function under consideration, such as ${I}$, $\Lambda$, $Q$ and ${\omega}$, and $\eta_{\text{fit}}$ represents the value estimated using the relation obtained between the quantities considered. 

\begin{table}
\centering
\footnotesize
\renewcommand{\arraystretch}{1.75} 
\begin{tabular}{!{\vrule width 1.1pt}c!{\vrule width 1.1pt}c|c|c|c|c!{\vrule width 1.1pt}
}
\cline{2-6}
\multicolumn{1}{c!{\vrule width 1.1pt}}{}&${\bf a_0}$ & ${\bf a_1}$ & ${\bf a_2}$ & ${\bf a_3}$ & ${\bf a_{4}}$\\
\specialrule{1.1pt}{1.1pt}{1.1pt}
$\Lambda-\bar{I}$&1.498 & 0.06254 & 0.02147 & -6.031$\times10^{-4}$ & 5.123$\times 10^{-6}$  \\
\hline
$\Lambda-\bar{q}$&0.1785&0.1011&0.0448&.-00385&1.068$\times10^{-4}$\\
\hline
$\bar{I}-\bar{q}$&-2.856&2.201&-0.00516&0.0873&0.0111\\
\hline
$\bar{I}-\bar{\omega}$ &0.2124&0.163&-0.1753&0.04818&-4.328$\times10^{-3}$\\
\hline
$\Lambda-\bar{\omega}$&0.1984&+0.001522&-0.005334&-5.076$\times10^{-4}$&-1.441$\times10^{-5}$\\
\hline
$\bar{q}-\bar{\omega}$&0.2124&-0.04373&-0.02271&8.201$\times10^{-3}$&-7.262$\times10^{-4}$\\
\hline
\multicolumn{1}{c!{\vrule width 1.1pt}}{} & \multicolumn{5}{c!{\vrule width 1.1pt}}{$\lambda_{BL}=0$} \\
\hline
$\Lambda-\bar{I}$&1.255 & 0.1097 & 0.01511 & -6.128$\times10^{-5}$ & -1.333$\times 10^{-5}$  \\
\hline
$\Lambda-\bar{q}$&0.4528&0.106&0.03595&.004524&-8.597$\times10^{-4}$\\
\hline
$\bar{I}-\bar{q}$&16.04&-33.75&25&-7.502&0.8078\\
\hline
$\bar{I}-\bar{\omega}$ &-0.2694&0.8141&-0.4743&0.1028&-7.239$\times 10^{-3}$\\
\hline
$\Lambda-\bar{\omega}$& 0.1961&-0.00463&-0.001939&9.862$\times10^{-4}$&7.751$\times10^{-5}$\\
\hline
$\bar{q}-\bar{\omega}$&0.1974&0.0129&-0.146&0.1286&0.07228\\
\hline
\multicolumn{1}{c!{\vrule width 1.1pt}}{}& \multicolumn{5}{c!{\vrule width 1.1pt}}{$\lambda_{BL}=2$} \\
\hline
$\Lambda-\bar{I}$&1.637 & 0.04998& 0.02102 & -5.199$\times10^{-4}$ &2.843$\times10^{-6}$  \\
\hline
$\Lambda-\bar{q}$&0.2967&0.05414&0.1184&-0.01541&6.589$\times10^{-4}$\\
\hline
$\bar{I}-\bar{q}$&-1.136 & 0.08961 & 0.8455 & -0.2305 &1.945$\times10^{-2}$  \\
\hline
$\bar{I}-\bar{\omega}$ &0.3631&-0.05468&-0.0598&0.02103&-1.951$\times 10^{-3}$\\
\hline
$\Lambda-\bar{\omega}$& 0.1901&+0.01108&-0.008253&8.194$\times 10^{-4}$&$-2.538\times10^{-5}$\\
\hline
$\bar{q}-\bar{\omega}$&0.2204&-0.06988&-0.01516&9.148$\times 10^{-3}$&-1.094$\times 10^{-4}$\\
\hline
\multicolumn{1}{c!{\vrule width 1.1pt}}{} & \multicolumn{5}{c!{\vrule width 1.1pt}}{$\lambda_{BL}=-2$} \\
\cline{2-6}
\end{tabular}
\caption{The correlation coefficients obtained between tidal deformability ($\Lambda$), scaled moment of inertia ($\bar{I}$), scaled quadrupole moment ($\bar{q}$), and scaled $f$-mode frequency ($\bar{\omega}$) for the dark energy stars with  anisotropic parameter values $\lambda_{BL}=-0,\;2,\;-2$. }\label{tab:1}
\end{table}

First, we proceed to study the $I$–Love relation for anisotropic dark energy stars by considering different values of the EoS parameters $A$ and $B$. To numerically compute the moment of inertia $I$, we employ the modified Hartle-Thorne slow rotation approximation for the anisotropic stars~\cite{Rahmansyah:2020gar}. This approximation has been applied successfully to study the rotational properties of anisotropic dark energy stars~\cite{Jyothilakshmi:2024zso}. The tidal deformability $\Lambda$ is calculated by numerically solving the differential equations given in Ref.~\cite{Biswas:2019gkw}, which is extended to include the anisotropic effects following the approach of Hinderer et al.~\cite{Hinderer2008,Hind2010}. For our analysis, we consider three values of the anisotropy parameter: $\lambda_{BL} = -2,\, 0,\, 2$.

\begin{figure}
\centering
\subfloat[$\lambda_{BL}=0$]{\includegraphics[width=\linewidth]{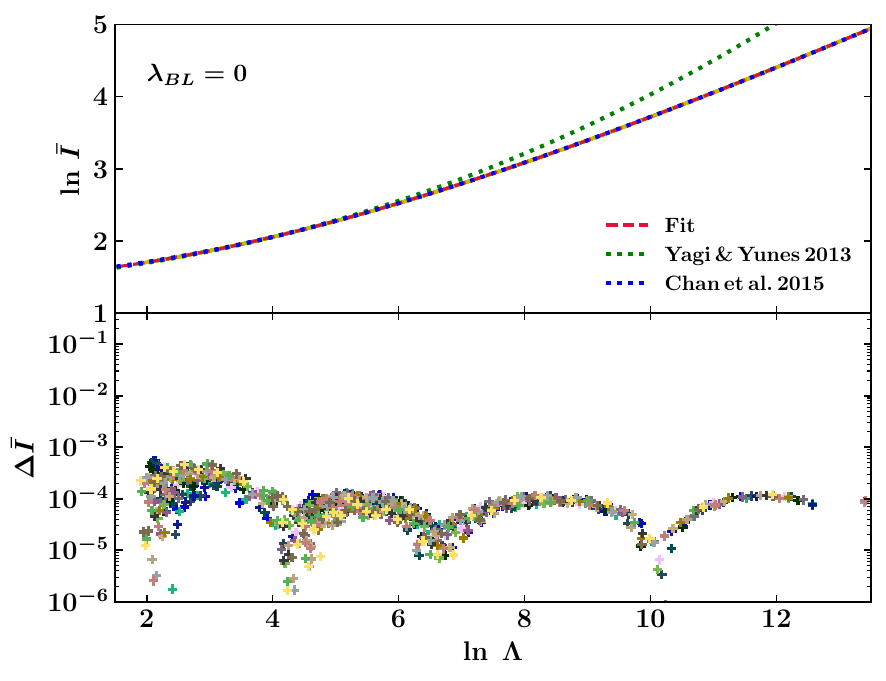}\label{fig:LI}}\\
\subfloat[$\lambda_{BL}=2$]{\includegraphics[width=\linewidth]{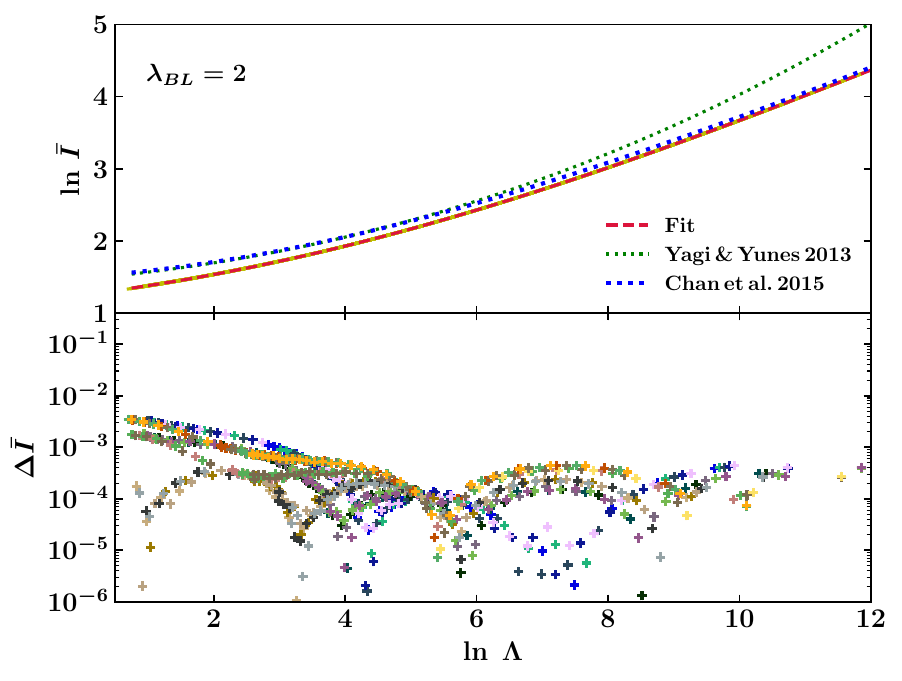}\label{fig:LI-2}}\\
\subfloat[$\lambda_{BL}=-2$]{\includegraphics[width=\linewidth]{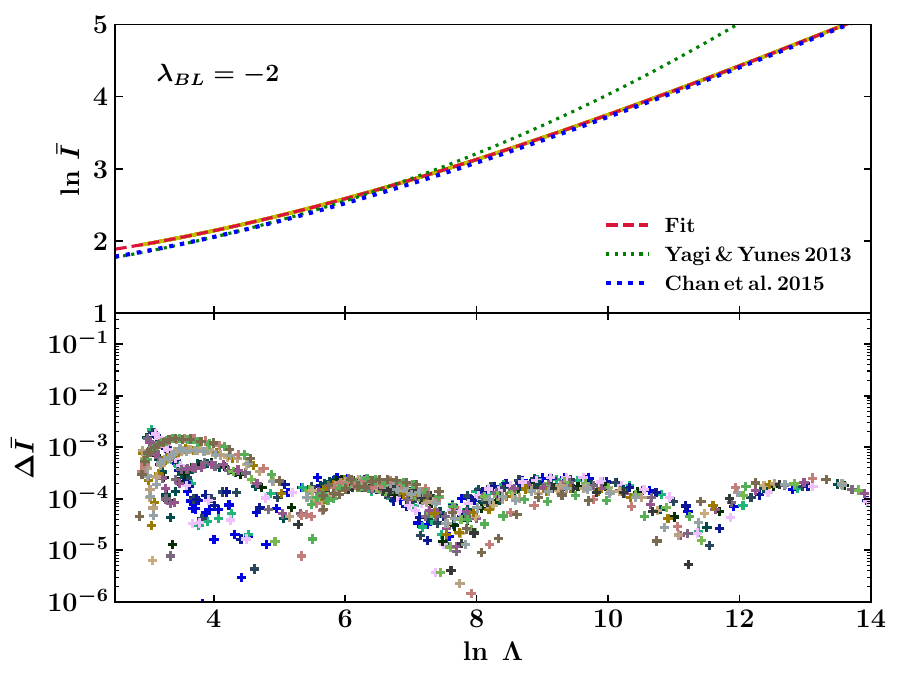}\label{fig:LI-1}}
\caption{The moment of inertia as a function of tidal deformability for different parameterizations of anisotropic dark energy stars.}
\label{IL}
\end{figure}

The relation between the normalized moment of inertia $\bar{I}$ and $\Lambda$ is written in terms of correlation coefficients $a_n$, expressed through a polynomial fit:
\begin{align}
    \ln\bar{I} = \sum_{n=0}^{4} a_n (\ln\Lambda)^n.
\end{align}
We first analyze the isotropic case with $\lambda_{BL} = 0$, for which the fitted relation is:
  $  \ln\bar{I} = 1.498  - 0.06245\, (\ln\Lambda) + 0.02147\, (\ln\Lambda)^2 - 6.031\times 10^{-4}\, (\ln\Lambda)^3 + 5.123\times 10^{-6}\,(\ln\Lambda)^4.$
This relation is shown in Fig.~\ref{fig:LI}, and we find that $\bar{I}$ varies almost universally with $\Lambda$. The maximum deviation of estimated values from the fit remains below 1\%. To explore the impact of anisotropy, we repeat the analysis for $\lambda_{BL} =  +2,\,-2$. The resulting relation for $\lambda_{BL} = 2$ is plotted in the upper panel of Fig.~\ref{fig:LI-1} along with the error calculated, which is shown in the lower panel. Similar to the isotropic case, the maximum deviation observed is below $1\%$ only. A corresponding analysis for $\lambda_{BL} = -2$ gives a relation (Fig.~\ref{fig:LI-2}), which also results in an error of $<1\%$. 
We also compared our results with earlier universal relations derived for other compact star models such as neutron star or quark star~\cite{Yagi:2013awa} and incompressible stars~\cite{Chan:2015iou}. We find that the relation obtained for dark energy stars are in agreement with those for other stellar matter compositions. 
The coefficients obtained for all the three fits are summarized in Table~\ref{tab:1}.
\begin{figure}
\centering
\subfloat[$\lambda_{BL}=0$]{\includegraphics[width=\linewidth]{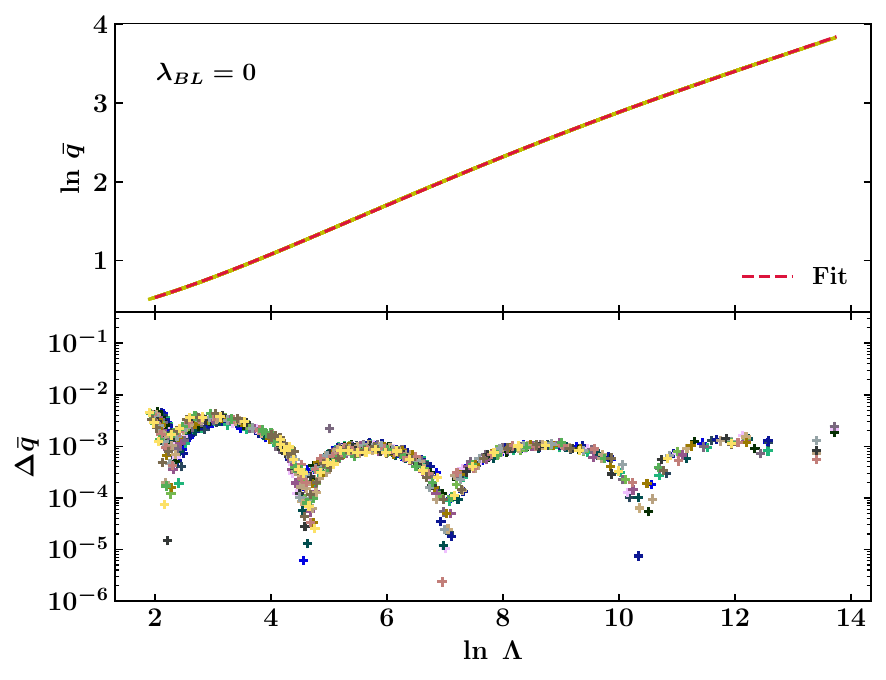}\label{fig:QL}}\\
\subfloat[$\lambda_{BL}=2$]{\includegraphics[width=\linewidth]{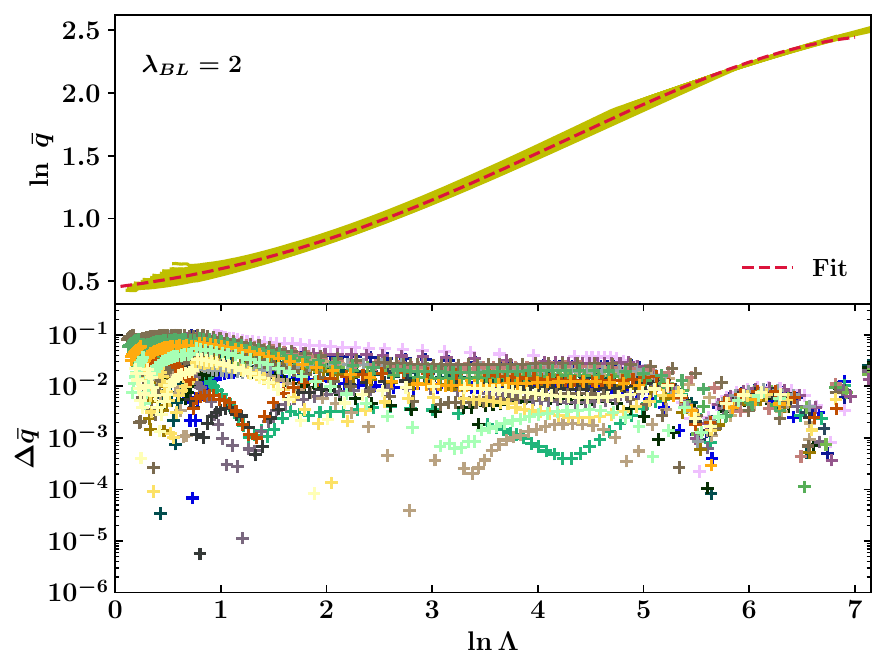}\label{fig:QL-2}}\\
\subfloat[$\lambda_{BL}=-2$]{\includegraphics[width=\linewidth]{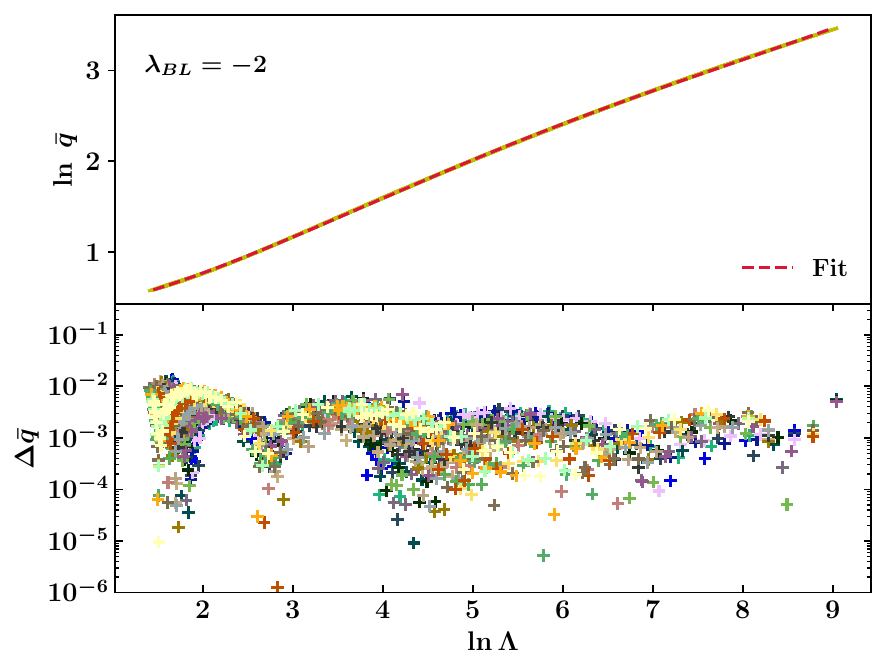}\label{fig:QL-1}}
\caption{The quadrupole moment as a function of tidal deformability for different parameterizations of anisotropic dark energy stars.}
\label{IL}
\end{figure}

\par
In the following, we derive a relation between normalized quadrupole moment $\bar{q}=Q/M^3$ and tidal deformability $\Lambda$ of the dark energy stars for $\lambda_{BL}=-2,\,0,\,2$. We obtain the relation in the form given below:
\begin{align}
    \ln \bar{q} = \sum_{n=0}^{4} a_n (\ln\Lambda)^n,
\end{align}
where $a_n$ is the fitting coefficient. The quadrupole moments are calculated by employing the mechanism described by Hartle and Thorne in Refs.~\cite{Hartle:1967he,Hartle:1968si} for $\lambda_{BL}=-2,\,0,\,2$. 
In the upper panel of Fig.~\ref{fig:QL}, we plot the $Q-\Lambda$ relation of the isotropic ($\lambda_{BL}=0$) dark energy star and we find that $\bar{q}$ varies universally with $\Lambda$. We also note that, as the value of $\Lambda$ tends to zero, that of $\bar{q}$ tends to unity. This relates our results with the no-hair theorem of black holes, which gives an estimate for $\bar{q}$, $\bar{I}$ and $\Lambda$ of black holes as $1$, $4$ and $0$ respectively~\cite{Yagi:2013awa}. We observe from the lower panel of Fig.~\ref{fig:QL}, that the maximum error obtained between the calculated fit and our result is around $\sim 1\%$. The $Q-\Lambda$ relation obtained for $\lambda_{BL}=2$ and $-2$ cases are shown in Fig.~\ref{fig:QL-2} and ~\ref{fig:QL-1}, respectively. We find that these relations follow a universal nature and the value of $\bar{q}$ tends to 1, as $\Lambda$ approaches 0.
The maximum deviation obtained for both $\lambda_{BL}=-2$ and $2$ are around $1-10\%$; and our fit gives a reasonable estimate of $\bar{q}$ and $\Lambda$, if either one of them is known. We have tabulated all the relevant coefficients for $Q-\Lambda$ relation in Table \ref{tab:1}.

\begin{figure}
\centering
\subfloat[$\lambda_{BL}=0$]{\includegraphics[width=\linewidth]{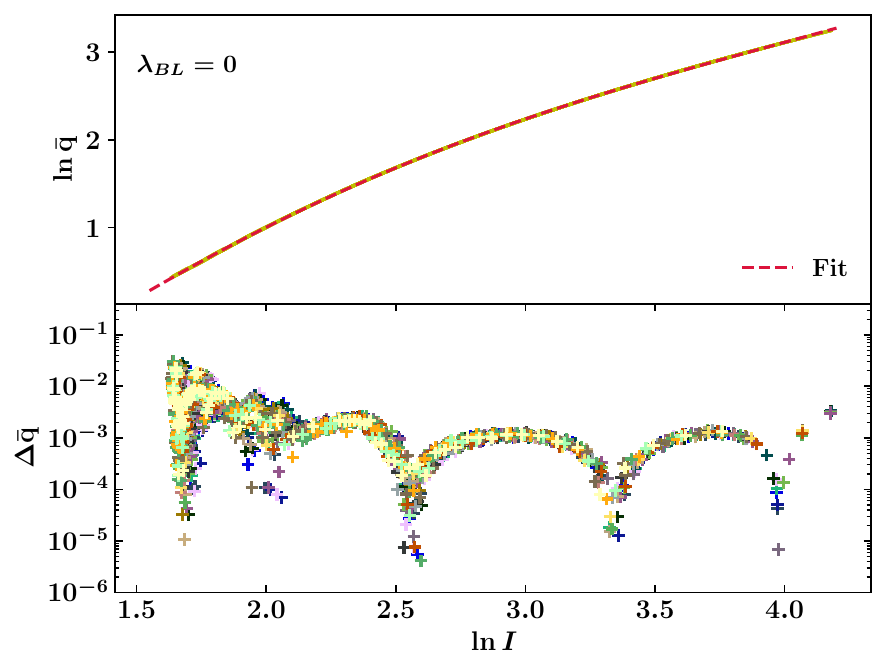}\label{fig:QI}}\\
\subfloat[$\lambda_{BL}=2$]{\includegraphics[width=\linewidth]{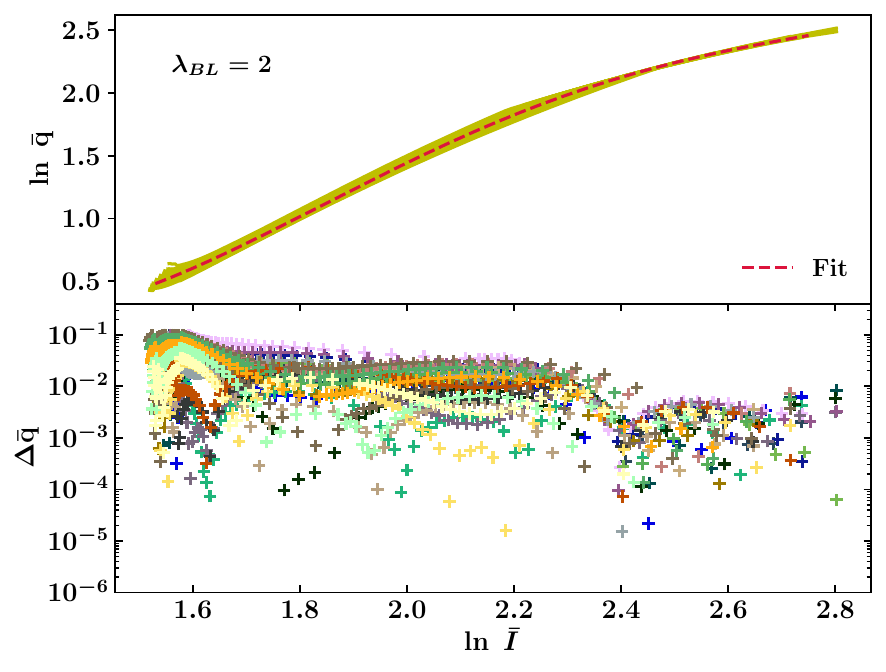}\label{fig:QI-2}}\\
\subfloat[$\lambda_{BL}=-2$]{\includegraphics[width=\linewidth]{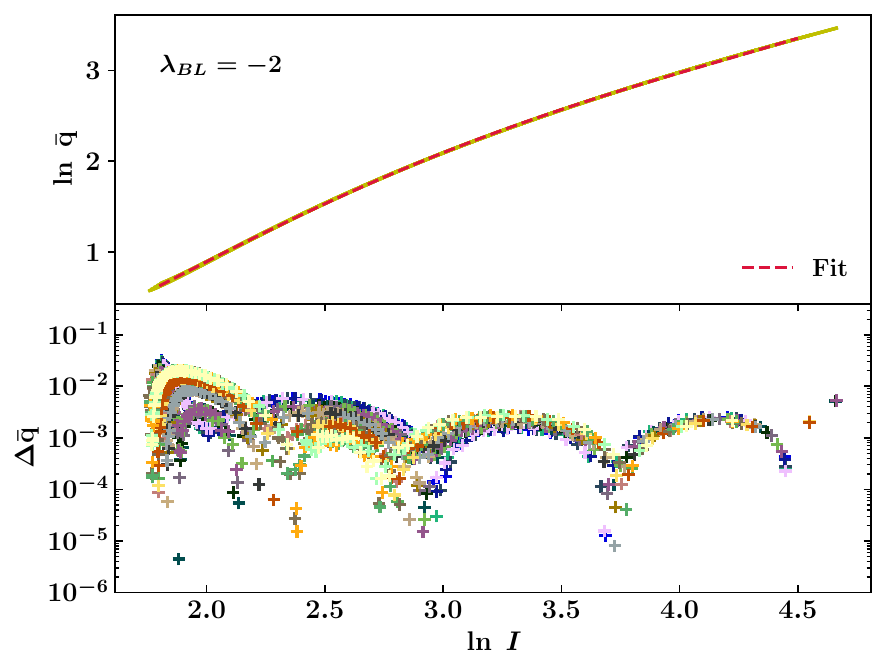}\label{fig:QI-1}}
\caption{The quadrupole moment as a function of moment of inertia for different parameterizations of anisotropic dark energy stars.}
\label{QI}
\end{figure}

\par 
Next, we do a similar analysis for normalized quadrupole moment $\bar{q}$ and moment of inertia $\bar{I}$. The relation is expressed in terms of the fitting coefficients $a_n$ as
\begin{align}
    \ln \bar{q} = \sum_{n=0}^{4} a_n (\ln I)^n.
\end{align}
Initially, we obtain the relation between $\bar{I}$ and $\bar{q}$ for the isotropic case ($\lambda_{BL}=0$) and is plotted in the upper panel of Fig.~\ref{fig:QI}. We observe that the $Q-I$ relation is universal for the $\lambda_{BL}=0$ case, as expected. Further, we find that the value of $\bar{I}$ approaches $4$, as $\bar{q}$ tends to $1$. Thus, these results too can be related to the no-hair theorem of black holes. In the lower panel of Fig.~\ref{fig:QI}, we have also shown the deviation of the numerical results from the fit. We observe an error percentage of $\leq 4\%$ for the isotropic case. 
Then, we proceed to obtain relations for anisotropic dark energy stars with values of $\lambda_{BL}= 2$ and $-2$ and they are shown in the upper panel of Fig.~\ref{fig:QI-2} and Fig.~\ref{fig:QI-1} respectively. For the $\lambda_{BL}=2$ case, we find a small spread for $\bar{q}$ and $\bar{I}$ spectra, indicating a slight deviation from the universal property generally seen for $\bar{q}$ and $\bar{I}$~\cite{Yagi:2013awa}. We also show the deviation of numerically obtained values from the fit, in the lower panel of Fig.~\ref{fig:QI-2} and Fig.~\ref{fig:QI-1}. We note that, the maximum error percentage for both $\lambda_{BL}= 2$ and $-2$ are $\sim 10\%$ and $\sim 2\%$, respectively.

\begin{figure}
\centering
\subfloat[$\lambda_{BL}=0$]{\includegraphics[width=\linewidth]{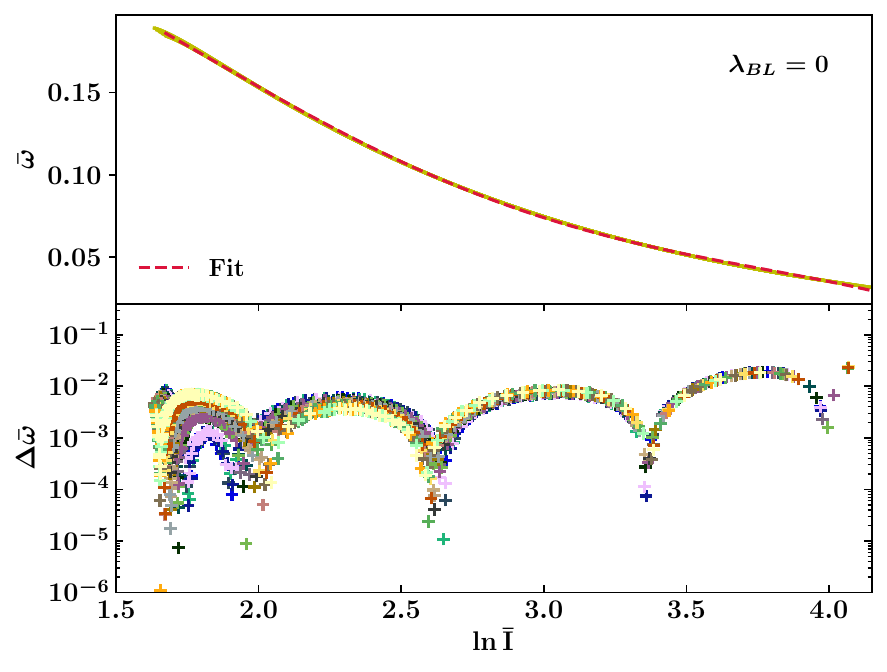}\label{fig:fI}}\\
\subfloat[$\lambda_{BL}=2$]{\includegraphics[width=\linewidth]{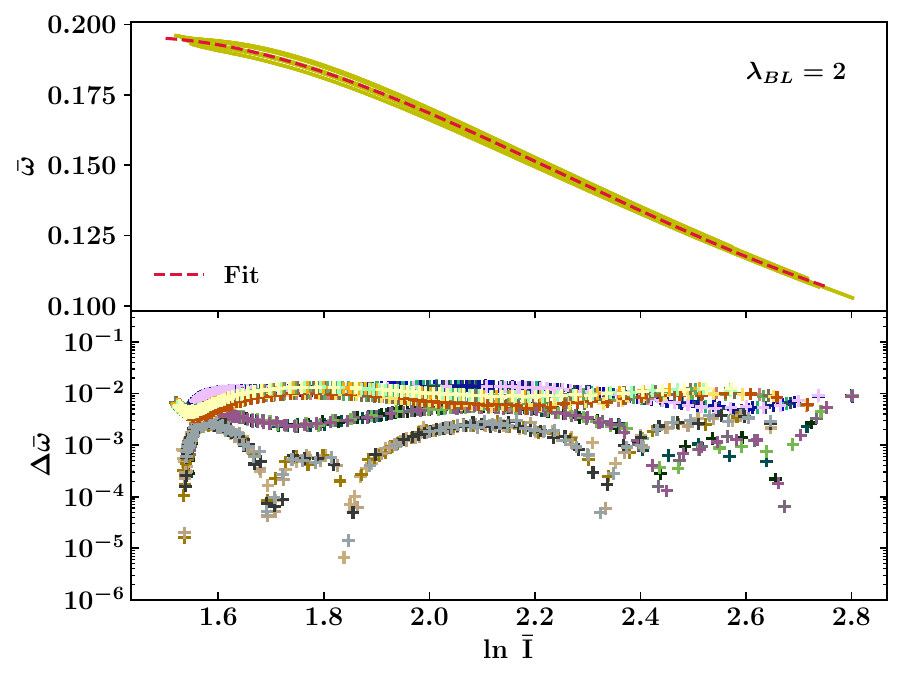}\label{fig:fI-2}}\\
\subfloat[$\lambda_{BL}=-2$]{\includegraphics[width=\linewidth]{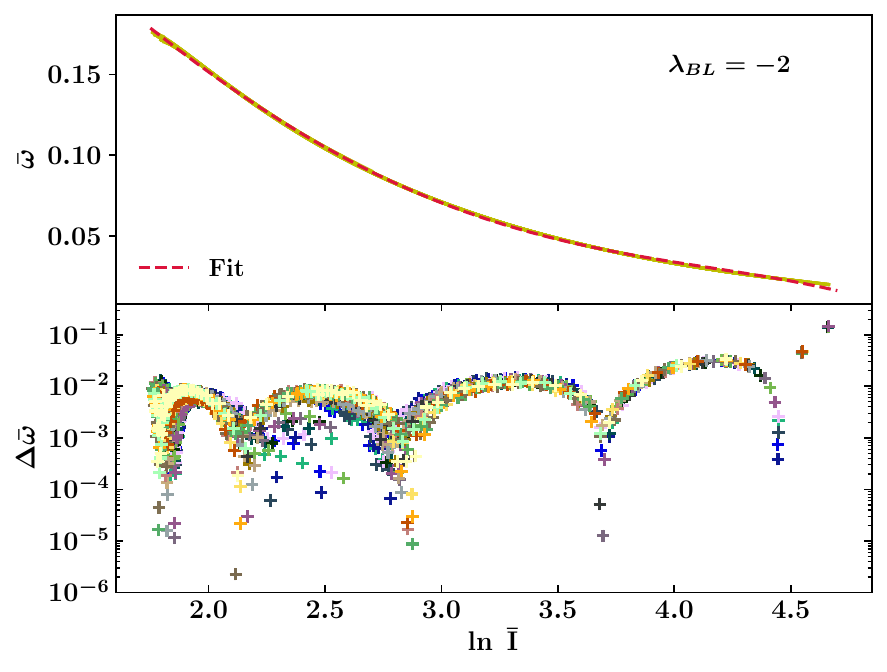}\label{fig:fI-1}}
\caption{The scaled angular velocity $\omega M$ as a function of moment of inertia $I$ for different parameterizations of anisotropic dark energy stars.}
\label{fI}
\end{figure}

\par
Further, we also investigate the correlations between the $f$-mode frequency and the quantities $I$, $\Lambda$ and $Q$. We employ the modified relativistic Cowling approximation~\cite{Doneva:2012rd,Curi:2022nnt,Jyothilakshmi:2024zqn} to obtain the $f$-mode frequencies of the anisotropic dark energy stars, by varying $\lambda_{BL}=-2,\;0,\;2$. In Fig.~\ref{fig:fI}, we study the correlation between mass scaled angular frequency ($\bar{\omega}=\omega M$) and normalized moment of inertia $\bar{I}$ for the isotropic case. The relation is expressed in terms of the coefficients $a_n$ as
\begin{align}
    \bar{\omega} = \sum_{n=0}^{4} a_n (\ln \bar{I})^n.
\end{align}
We find that $\bar{\omega}$ varies universally with $\bar{I}$ and the relation obtained is: $\bar{\omega}= 0.2124 + 0.163\,(\ln \bar{I}) - 0.1753\, (\ln \bar{I})^2 + 0.04818\,(\ln \bar{I})^3 - 4.328\times10^{-3}\,(\ln \bar{I})^4$. We also plot the deviation of numerical fit from the estimated values, in the lower panel of Fig.~\ref{fig:fI}. We observe that the maximum deviation is about $\sim 4\%$ for the isotropic case. 
We now proceed to obtain a similar relation for the anisotropic case with $\lambda_{BL}=2$ and $-2$. The relation obtained between $\bar{\omega}$ and $\bar{I}$ with $\lambda_{BL}=2$ is: $\bar{\omega}= -0.2694 + 0.8141\,(\ln \bar{I}) - 0.4743\, (\ln \bar{I})^2 + 0.1028\,(\ln \bar{I})^3 - 7.239\times10^{-3}\,(\ln \bar{I})^4$ and is plotted in Fig.~\ref{fig:fI-2}. Though, $\bar{\omega}$ varies universally with $\bar{I}$ for $\lambda_{BL}=0$ case, it shows a slight spread in the spectra for $\lambda_{BL}=2$. 
We obtain the relation for $\lambda_{BL}=-2$ as $\bar{\omega}= 0.3631 - 0.05468 \,(\ln \bar{I}) - 0.0598\, (\ln \bar{I})^2 + 0.02103\,(\ln \bar{I})^3 - 1.951\times10^{-3}\,(\ln \bar{I})^4$; and is depicted in Fig.~\ref{fig:fI-1}. We note that, the maximum deviation from the estimated values is about $\sim 1\%$ ($\lambda_{BL}=2$) and $\sim 4\%$ ($\lambda_{BL}=-2$).

\begin{figure}
\centering
\subfloat[$\lambda_{BL}=0$]{\includegraphics[width=\linewidth]{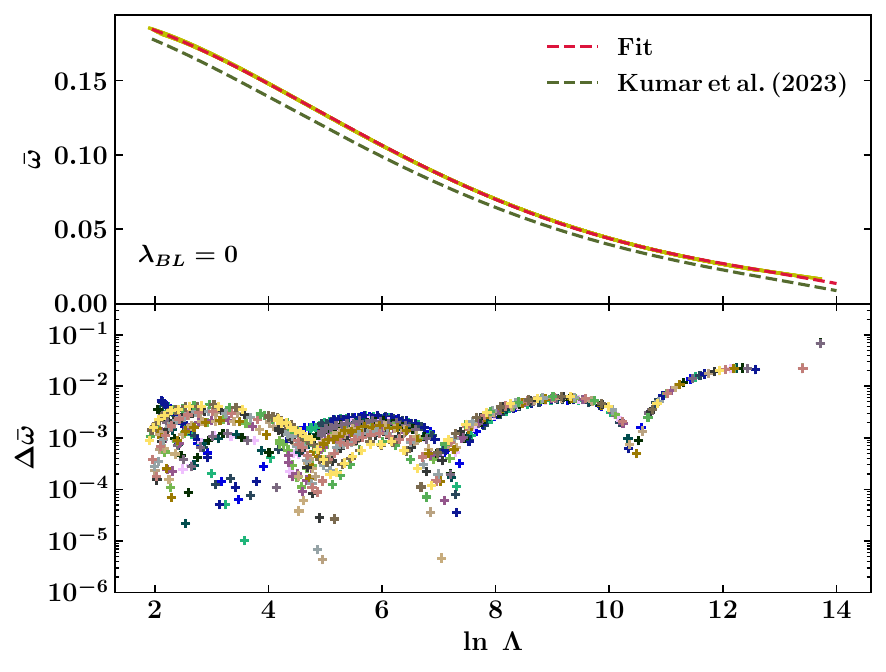}\label{fig:fL}}\\
\subfloat[$\lambda_{BL}=2$]{\includegraphics[width=\linewidth]{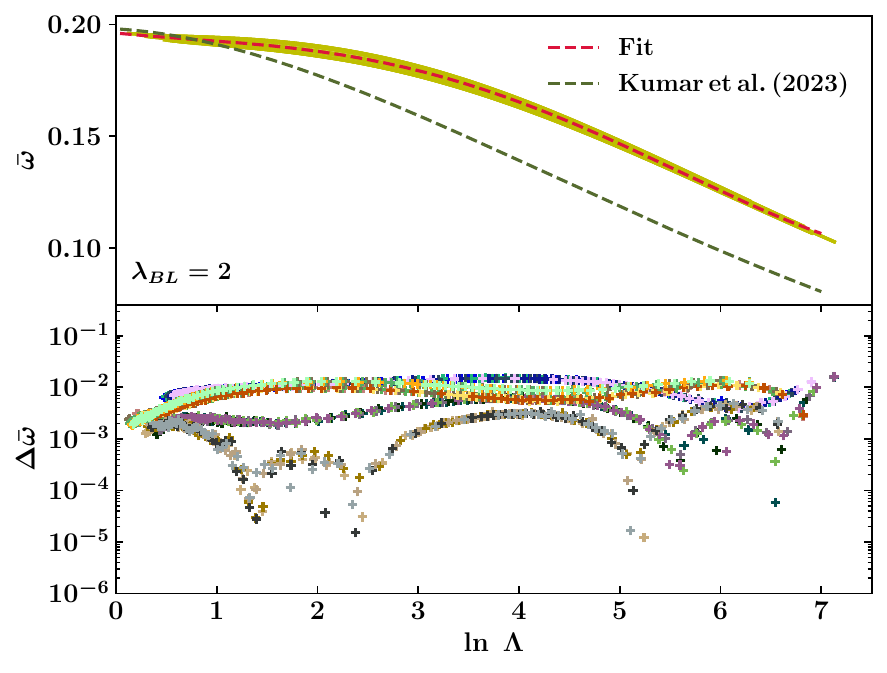}\label{fig:fL-2}}\\
\subfloat[$\lambda_{BL}=-2$]{\includegraphics[width=\linewidth]{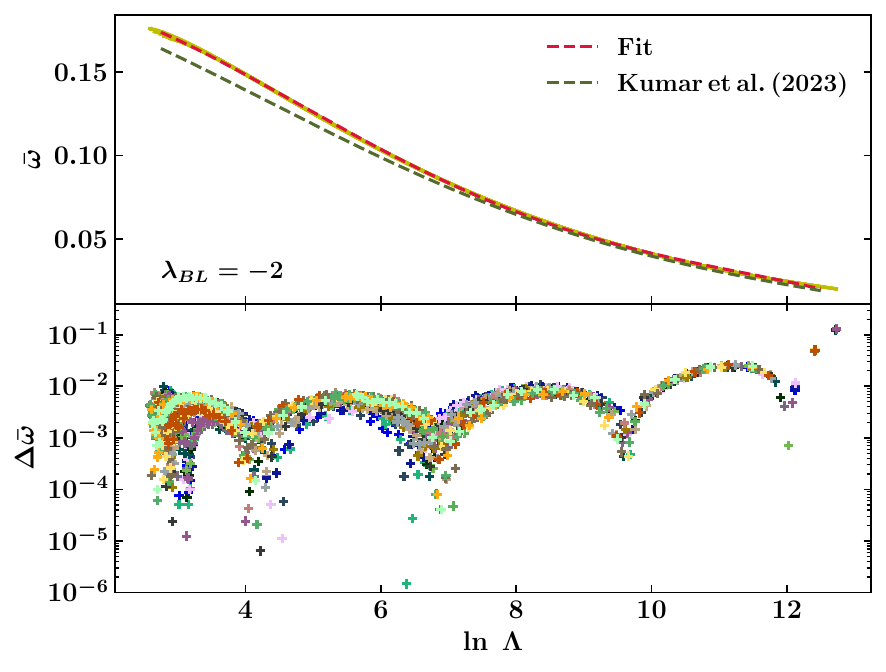}\label{fig:fL-1}}
\caption{The scaled angular velocity $\omega M$ as a function of tidal deformability $\Lambda$ for different parameterizations of anisotropic dark energy stars.}
\label{fL}
\end{figure}

\begin{figure}
\centering
\subfloat[$\lambda_{BL}=0$]{\includegraphics[width=\linewidth]{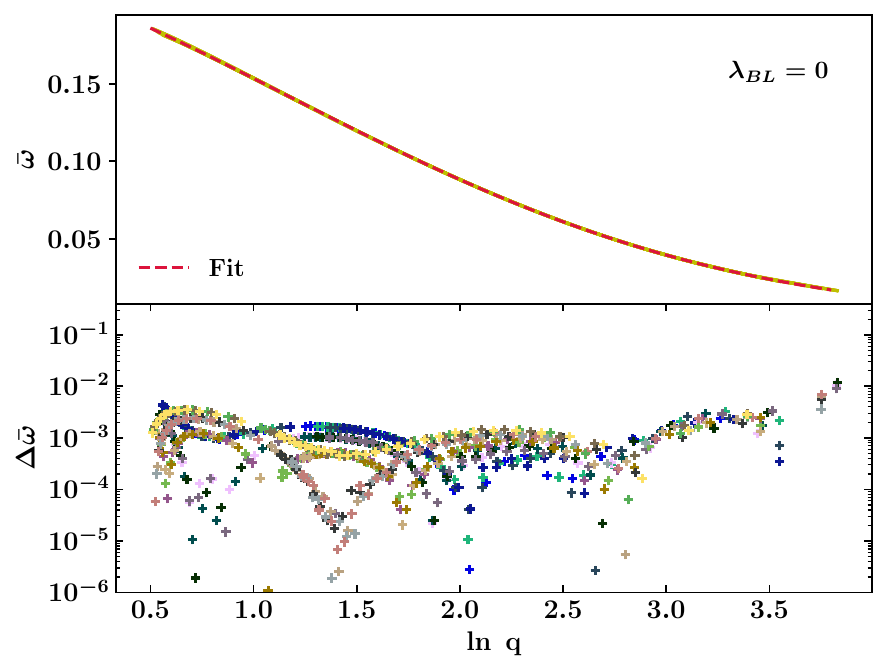}\label{fig:fq}}\\
\subfloat[$\lambda_{BL}=2$]{\includegraphics[width=\linewidth]{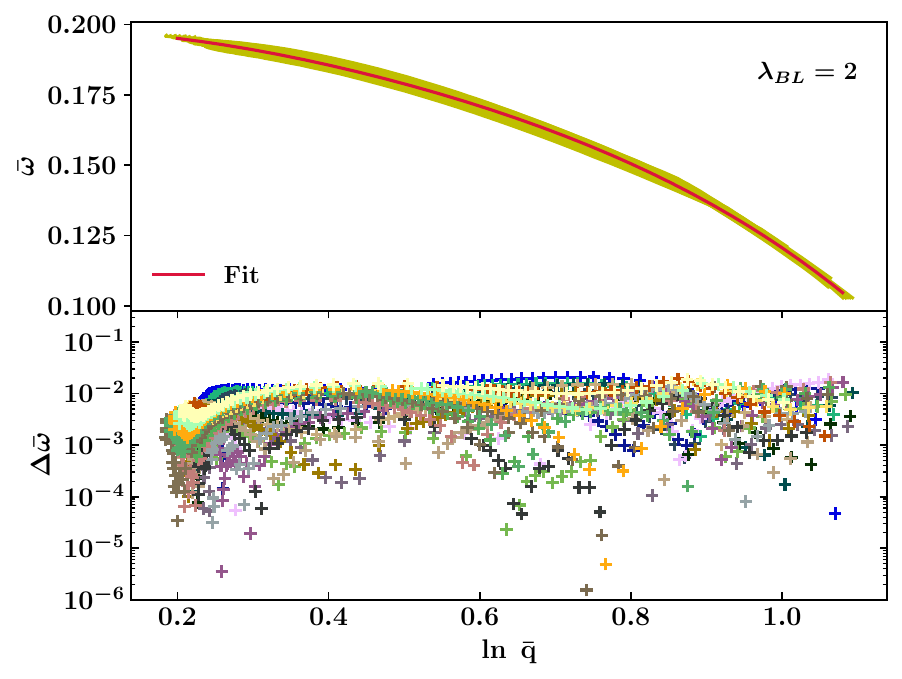}\label{fig:fq-2}}\\
\subfloat[$\lambda_{BL}=-2$]{\includegraphics[width=\linewidth]{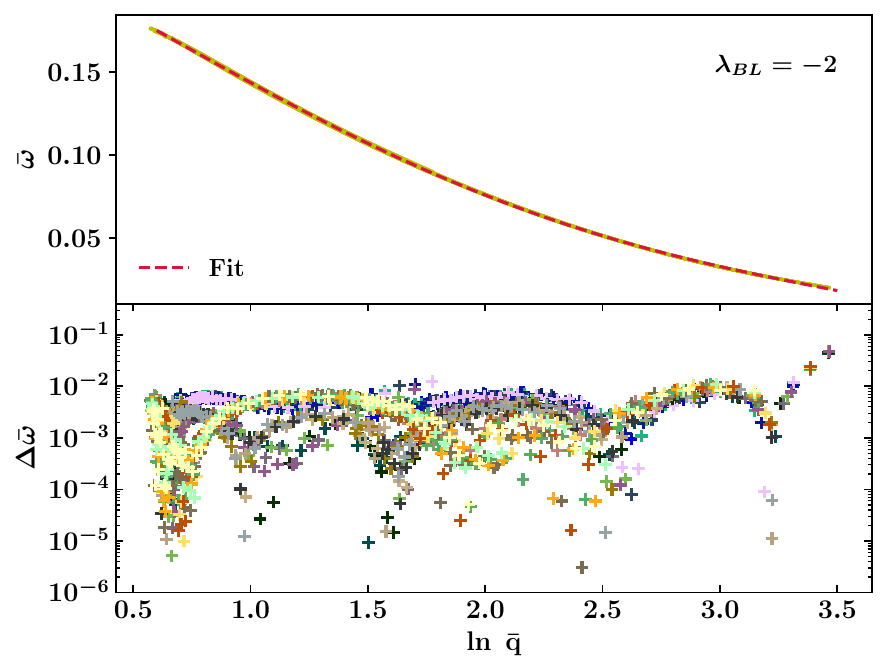}\label{fig:fq-1}}
\caption{The scaled angular velocity $\omega M$ as a function of quadrupole moment $\bar{q}$ for different parameterizations of anisotropic dark energy stars.}
\label{fq}
\end{figure}

\par 
The scaled angular frequency $\bar{\omega}$, is plotted against the dimensionless tidal deformability $\Lambda$, by varying the anisotropic parameter $\lambda_{BL}= 0,\;2,\;-2$, in Figs.~\ref{fig:fL}, \ref{fig:fL-2}, and \ref{fig:fL-1}, respectively. 
The relation between $\bar{\omega}$ and $\Lambda$ can be given as
\begin{align}
    \bar{\omega} = \sum_{n=0}^{4} a_n (\ln \Lambda)^n.
\end{align}
We find that $\bar{\omega}$ varies universally with $\Lambda$, for the isotropic case ($\lambda_{BL}=0$), as plotted in the upper panel of Fig.~\ref{fig:fL}. For comparison, we have also shown the relation obtained between $\bar{\omega}$ and $\Lambda$ in Ref.~\cite{Kumar:2023ojk} for compact stars with exotic degrees of freedom. We observe that, both are in close agreement with each other. In the lower panel of Fig.~\ref{fig:fL}, we show the deviation from the numerical fit. The maximum error percentage obtained is only $\sim 2\%$.   
Next, we obtain the numerical fits between scaled angular frequency $\bar{\omega}$ and dimensionless tidal deformability $\Lambda$ in presence of anisotropy with $\lambda_{BL}=-2,\;2$. As we see from the upper panel of Fig.~\ref{fig:fL-2}, $\bar{\omega}$ and $\Lambda$ show a small spread in the spectra, indicating a slight deviation from the universal nature observed in the isotropic scenario. We also plot the relations obtained for isotropic cases, considered in earlier works, along with that for the isotropic dark energy stars. We find a slight deviation from the isotropic case as expected. Finally, we also plot the relation between $\bar{\omega}$ and $\Lambda$ for $\lambda_{BL}=-2$ case in Fig.~\ref{fig:fL-1}. We note that, unlike the $\lambda_{BL}=2$ case, $\bar{\omega }$ varies universally with $\Lambda$, for $\lambda_{BL}=-2$. Like the previous cases, we have also shown relations obtained for other compact stars in Fig.~\ref{fig:fL-1}. For both $\lambda_{BL}=-2$ and $2$, the deviation from the estimated values are found to be less than $4\%$. 

\begin{figure}
    \centering
    \includegraphics[width=\linewidth,height=2.5in]{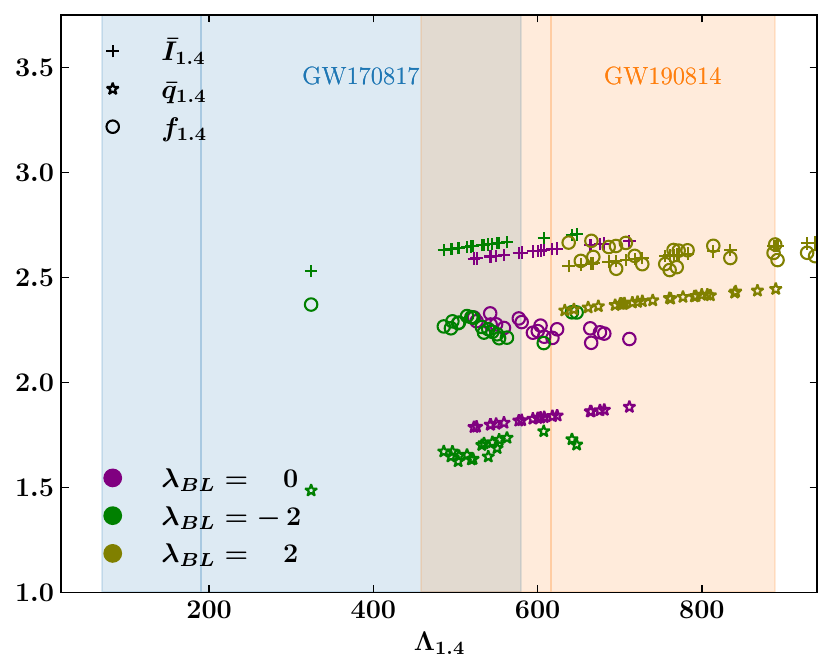}
    \caption{Observables such as moment of inertia $\bar{I}_{1.4}$, quadrupole moment $\bar{q}_{1.4}$, and the $f$-mode frequency $f_{1.4}$ (in kHz) of anisotropic dark energy stars (with $M=1.4M_\odot$) as function of tidal deformability $\Lambda_{1.4}$.  
    The color bands represent the $\Lambda_{1.4}$ constraints from GW170817 and GW190814 events.}
    \label{fig:GW}
\end{figure}
\begin{table}[htbp]
\centering
\renewcommand{\arraystretch}{1.75} 
\begin{tabular}{!{\vrule width 1.1pt}c!{\vrule width 1.1pt} c !{\vrule width 1.1pt}c !{\vrule width 1.1pt} c !{\vrule width 1.1pt} c !{\vrule width 1.1pt}}
\cline{3-5}
\multicolumn{2}{c!{\vrule width 1.1pt}}{}&$\lambda_{BL}=0$ & $\lambda_{BL}=-2$ & $\lambda_{BL}=2$ \\
\specialrule{1.1pt}{1.pt}{1.pt}
$\bar{I}_{1.4}$ &GW170817 & $10.32^{+3.39}_{-2.10}$ & $11.08^{+3.46}_{-2.15}$ & $9.28^{+3.24}_{-2.00}$ \\
\cline{2-5}
& GW190814 & $13.94^{+1.47}_{-1.06}$ & $14.78^{+1.50}_{-1.08}$ & $12.73^{+1.41}_{-1.01}$ \\
\specialrule{1.1pt}{1.pt}{1.pt}
$\bar{q}_{1.4}$ &  GW170817 & $4.337^{+1.828}_{-1.159}$ & $8.280^{+4.464}_{-2.824}$ & $7.394^{+2.978}_{-2.346}$ \\
\cline{2-5}
& GW190814 & $6.282^{+0.762}_{-0.558}$ & $13.029^{+1.844}_{-1.357}$ & $10.511^{+0.722}_{-0.726}$ \\
\specialrule{1.1pt}{1.pt}{1.pt}
$f_{1.4}$  & GW170817 & $2.806^{+0.481}_{-0.521}$ & $2.767^{+0.520}_{-0.557}$ & $3.257^{+0.450}_{-0.537}$  \\
\cline{2-5}
(kHz) & GW190814 & $2.257^{+0.134}_{-0.161}$ & $2.182^{+0.142}_{-0.168}$ & $2.692^{+0.137}_{-0.157}$ \\
\specialrule{1.1pt}{1.pt}{1.pt}
\end{tabular}
\caption{Constraints from GW170817 and GW190814 on observables of dark energy stars with different anisotropy parameters $\lambda_{BL}$.}
\label{tab:observ}
\end{table}

\begin{figure}
    \centering
\includegraphics[width=\linewidth,height = 2.5in]{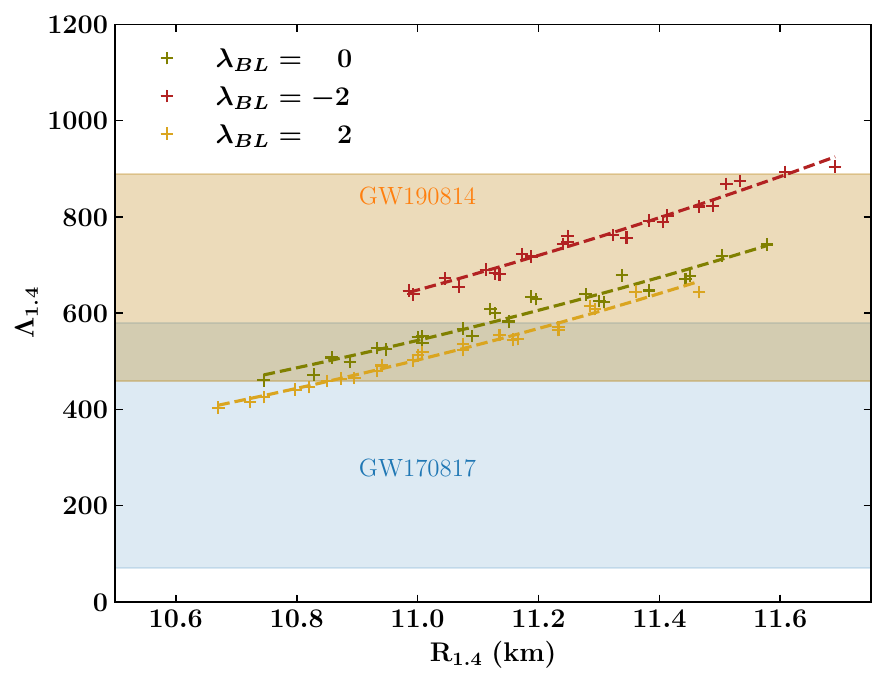}
    \caption{Canonical tidal deformability $\Lambda_{1.4}$ versus radius $R_{1.4}$ for different parameterizations of anisotropic dark energy stars. The dashed lines represent the exponential fit of the form $\Lambda_{1.4} = aR_{1.4}^b$ for 
    different $\lambda_{BL}$ values.  }
    \label{fig:GW-1}
\end{figure}
\begin{figure}
\centering
\subfloat[$\lambda_{BL}=0$]{\includegraphics[width=\linewidth]{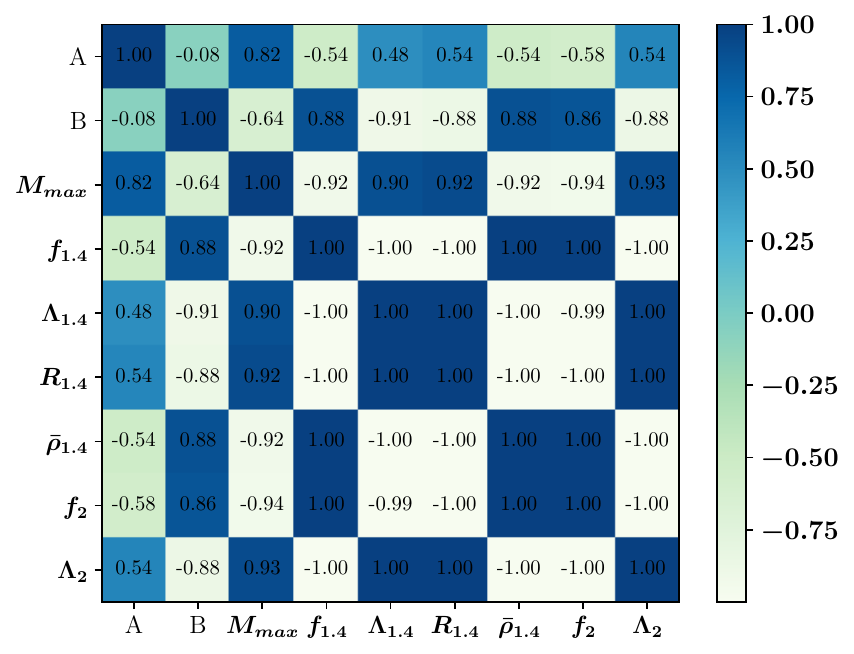}\label{fig:heat}}\\
\subfloat[$\lambda_{BL}=2$]{\includegraphics[width=\linewidth]{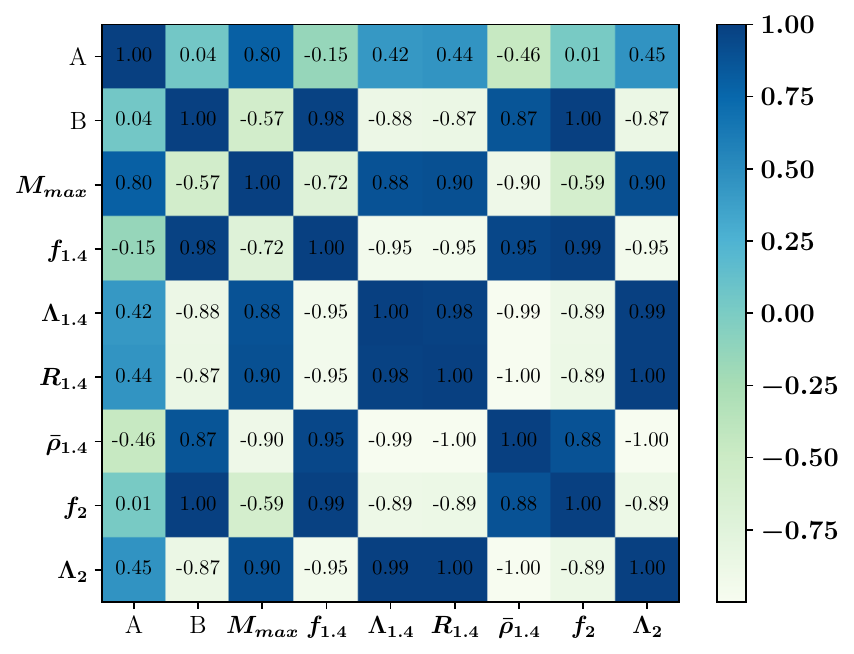}\label{fig:heat-2}}\\
\subfloat[$\lambda_{BL}=-2$]{\includegraphics[width=\linewidth]{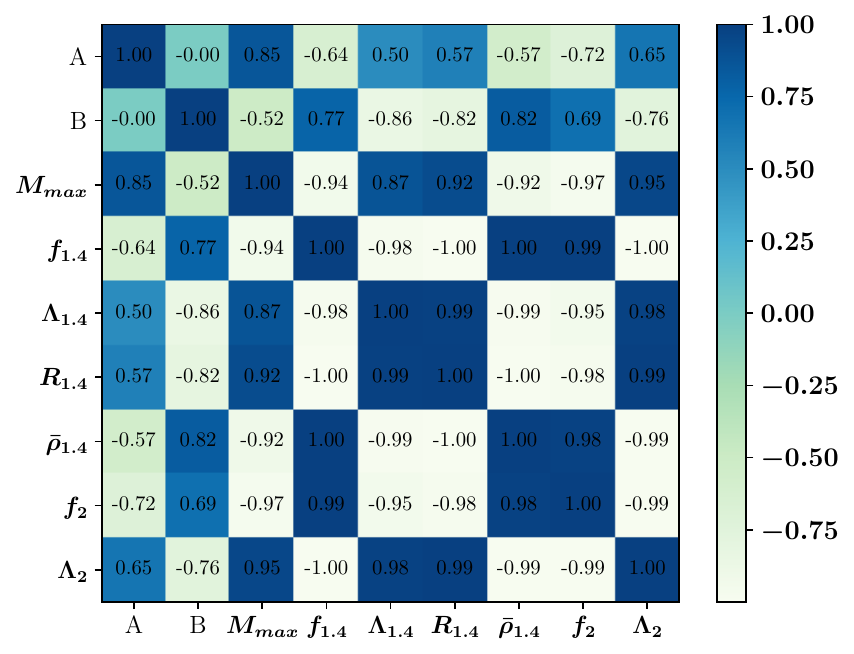}\label{fig:heat-1}}
\caption{Heat map showing the correlation of model parameters and stellar properties of anisotropic dark energy stars with $\lambda_{BL}=-2,\,0$ and $+2$.}
\label{IL}
\end{figure}

Further, the scaled angular frequency $\bar{\omega}$ is plotted against the dimensionless quadrupole moment $\bar{q}$, for the anisotropic parameters $\lambda_{BL} = 0,\; 2,\; -2$ in Fig.~\ref{fq}. The relationship between $\bar{\omega}$ and $\bar{q}$ is expressed as:
\begin{align} 
\bar{\omega} = \sum_{n=0}^{4} a_n (\ln \bar{q})^n; 
\end{align}
where $a_n$ are the correlation coefficients. For the isotropic case ($\lambda_{BL} = 0$), shown in the upper panel of Fig.~\ref{fig:fq}, we observe that $\bar{\omega}$ varies universally with $\bar{q}$. The lower panel of Fig.~\ref{fig:fq} illustrates the deviation of the numerically obtained values from the fitted relation for $\lambda_{BL} = 0$, with a maximum observed error percentage of $1\%$.
Next, we derive the numerical fit between $\bar{\omega}$ and $\bar{q}$, for the  anisotropic dark energy stars with $\lambda_{BL} = 2$. As shown in the upper panel of Fig.~\ref{fig:fq-2}, a slight spread in the spectra is observed, indicating a deviation from the universal behavior. 
Also, for the $\lambda_{BL} = -2$ case, shown in Fig.~\ref{fig:fq-1}, we find that $\bar{\omega}$ varies universally with $\bar{q}$, similar to the isotropic case. Relations for other compact stars are also included in Fig.~\ref{fig:fq-1} for comparison. For both $\lambda_{BL} = -2$ and $\lambda_{BL} = 2$, we observe that, the deviations from the estimated values remain below $4\%$.

The LIGO–Virgo Collaboration has reported observational constraints on the canonical tidal deformability (for a compact star with $M=1.4M_\odot$), namely $\Lambda_{1.4}= 190^{+390}_{-120}$ from  GW170817~\cite{LIGOScientific:2018cki} and $\Lambda_{1.4}=616^{+273}_{-158}$ from  GW190814~\cite{LIGOScientific:2020zkf}. With the universal relations obtained above and these observational limits, we proceed to calculate the theoretical limits for the canonical moment of inertia ($\bar{I}_{1.4}$), quadrupole moment ($\bar{q}_{1.4}$), and the $f$-mode frequency ($f_{1.4}$) of dark energy stars. 
In Fig.~\ref{fig:GW}, we show $\bar{I}_{1.4}$, $\bar{q}_{1.4}$, and $f_{1.4}$ as functions of $\Lambda_{1.4}$, for different values of the anisotropy parameter $\lambda_{BL}$. The constraints imposed on these quantities from the GW events are calculated from the universal relations, and are tabulated in Tab.~\ref{tab:observ}. 
For the $\lambda_{BL}=0$ case, the constraints obtained for $f_{1.4}$ are $2.806^{+0.481}_{-0.521}$ and $2.257^{+0.134}_{-0.161}$ kHz from GW170817 and GW190814, respectively. 
In Ref.~\cite{Mohanty:2023hha}, the authors have computed limits on $f_{1.4}$ for anisotropic neutron stars employing the Cowling approximation. For the isotropic case, they obtained the constraints from GW170817 and GW190814 as $f_{1.4} = 2.606^{0.457}_{-0.484}$ and $2.097^{+0.124}_{-0.149}$ kHz, respectively, which is more closer to the limit obtained for dark energy stars.

\par 
Next, we plot the canonical tidal deformability as a function of the radius in Fig.~\ref{fig:GW-1}. 
We find that $\Lambda_{1.4}$ values obtained for each EoS profile lie well within the range imposed by the above mentioned GW events.  
We also compute the exponential fit relations of the form $\Lambda_{1.4} = a R_{1.4}^n$, for the $\lambda_{BL}=0,\,2,\,-2$ cases, and are given by $\Lambda_{1.4} = (2.7R_{_{1.4}}^{6.046},\,5.3 R_{_{1.4}}^{5.843},\,0.4R_{_{1.4}}^{6.797})\times 10^{-4}$, respectively (see the dashed lines in Fig~\ref{fig:GW-1}). The corresponding correlation coefficients are found to be $0.978$, $0.987$ and $0.991$, respectively. Thus, we get that for $|\lambda_{BL}|>0$, the resulting correlation is stronger compared to the isotropic case ($\lambda_{BL}=0$).  
Further, using the constraints on tidal deformability from the GW events, we evaluate bounds on $R_{1.4}$ for dark energy star. We find that the limits imposed on $R_{1.4}$ from GW170817 are $R_{1.4} = 9.27^{+1.88}_{-1.40},\;8.93^{1.88}_{1.40}$ and $9.6^{+1.71}_{-1.31}$ km, respectively, for $\lambda_{BL} = 0,\,2,\,-2$. Similarly, for GW190814, we get $R_{1.4} = 11.26^{+0.71}_{-0.54},\; 10.92^{+0.71}_{-0.54}$ and $11.41^{+0.64}_{-0.48}$ km, respectively. This result is consistent with previously reported constraint for isotropic neutron stars, $R_{1.4} = 10.74^{+1.84}_{-1.36}$~\cite{Das:2022ell}.

\par
 Linear correlations between any two quantities are quantified using Pearson’s correlation coefficient, defined as~\cite{Benesty2009}
\begin{equation}
r(x,y) = \frac{\sigma_{xy}}{\sqrt{\sigma_{xx} \sigma_{yy}}}; \label{eq:P-coef}
\end{equation}
where, the covariance $\sigma_{xy}$ is expressed as
\begin{equation}
\sigma_{xy} = \frac{1}{N} \sum_i x_i y_i - \left( \frac{1}{N} \sum_i x_i \right) \left( \frac{1}{N} \sum_i y_i \right).\label{eq:sigma}
\end{equation}
Here, the index $i$ runs over the total number of models $N$~\cite{Brandt:1976zc}.  Also, $x_i$ and $y_i$ represent the values of two different stellar properties corresponding to a fixed mass, calculated from a set of theoretical models. A correlation coefficient with a magnitude close to the unity signifies a strong linear relationship between the considered pair of quantities. 
We now proceed to perform Pearson's correlation analysis for our anisotropic stellar system, by calculating the coefficients between various quantities, including the model parameters ($A$ and $B$), the maximum stellar mass ($M_{\rm max}$), and canonical properties, namely the mode frequency ($f_{1.4}$), tidal deformability ($\Lambda_{1.4}$), radius ($R_{1.4}$), and average density ($\bar{\rho}_{1.4}$). Additionally, correlations are evaluated for the mode frequency ($f_2$) and tidal deformability ($\Lambda_2$) at $M = 2\,M_\odot$, as well.  
\par 
The correlation heatmap shown in Fig.~\ref{fig:heat}, visualizes the relationships between EoS parameters ($A$, $B$) and stellar observables, including the maximum mass $M_{\text{max}}$, fundamental oscillation frequencies $f_{1.4}$ and $f_2$, tidal deformabilities ($\Lambda_{1.4}$, $\Lambda_2$), radius $R_{1.4}$, and mean density $\bar{\rho}_{1.4}$ of the isotropic dark energy stars ($\lambda_{BL}=0$).
 We find that, the EoS parameter $B$ exhibits a strong positive correlation with the fundamental frequencies ($f_{1.4}$, $f_2$) and density $\rho_{1.4}$. We also note that, $B$ is negatively correlated with radius ($R_{1.4}$) and tidal deformabilities ($\Lambda_{1.4}$, $\Lambda_2$). We observe that, $A$ shows a weaker correlation with other quantities and also exhibits an opposite trend compared to that of $B$. The maximum mass shows a strong correlation with tidal deformability (positive) and $f$-mode frequency (negative). 
 
 Further, a similar correlation map is obtained for the dark energy stars with anisotropy parameter $\lambda_{BL}=2,\,-2$. From Fig.~\ref{fig:heat-2}, with $\lambda_{BL}=2$, we observe a stronger positive correlation between the model parameter $B$ and $f$-mode frequencies, compared to the isotropic case. 
 However, $B$ is related to the radius and the tidal deformabilities negatively, same as that observed for the isotropic case. Also, we note that the maximum mass shows a less negative correlation with $f$-mode frequencies, compared to the $\lambda_{BL}=0$ case. We find that, the model parameter $A$, however has a very weak correlation with all the stellar attributes, except the maximum mass considered here. 
 Finally, in Fig.~\ref{fig:heat-1}, we show the correlation map for the anisotropic dark energy stars with $\lambda_{BL}=-2$. We estimate that, compared to the isotropic case, the correlation between $B$ and the mode frequencies are weaker. However, $B$ shows a negative correlation with the tidal defomability and the radius, similar to the previous cases. 
 Further, we find that, the EoS parameter $A$ exhibits a weaker correlation, with opposite trends, compared to the parameter $B$.

\section{Summary}\label{sec:summ}

In this Letter, we have investigated the universal relations among the moment of inertia, quadrupole moment, tidal deformability, and the $f$-mode frequency of anisotropic dark energy stars modeled with a modified Chaplygin equation of state and Bowers–Liang anisotropy. 
We have employed the modified Hartle–Thorne slow rotation and relativistic Cowling approximations to obtain the global stellar properties and the $f$-mode frequency. 
We report that all the relations show universal nature, with deviations only between $\sim 1-10\%$, though positive anisotropy induces a mild spread in the quadrupole and $f$-mode spectra.
\par 

 Using the inferred tidal deformability bounds from GW170817 and GW190814, we derived astrophysical constraints on canonical stellar properties: for a positive anisotropy strength, the $f$-mode frequency lies within $3.26^{+0.450}_{-0.537}$ kHz and $2.69^{+0.137}_{-0.157}$ kHz, and the stellar radius is constrained to $8.93^{1.88}_{1.40}$ km and $10.92^{+0.71}_{-0.54}$ km, respectively. Similarly, the dimensionless moment of inertia and quadrupole moment are constrained to be $9.28^{+3.24}_{-2.00}$ and $12.73^{+1.41}_{-1.01}$, and $7.39^{+2.978}_{-2.346}$ and $10.51^{+0.722}_{-0.726}$, respectively.
 
\par 

Finally, by applying Pearson correlation analysis for the first time to anisotropic compact stars, we demonstrate that the Chaplygin EoS parameter $B$ correlates positively with the $f$-mode frequency and negatively with radius and tidal deformability, while anisotropy modulates the correlation strength.  These results establish that universal relations extend to anisotropic dark energy stars and can be directly confronted with current and future gravitational-wave observations.

\par
\bibliography{ref}
\end{document}